\documentclass[fleqn,twoside]{article}
\usepackage{espcrc2}
\usepackage{verbatim}

\usepackage{psfrag}
\usepackage{amsfonts}
\usepackage[dvips,dvipdfm]{graphicx}
\usepackage[figuresright]{rotating}

\newcommand{\AmS}{{\protect\the\textfont2
  A\kern-.1667em\lower.5ex\hbox{M}\kern-.125emS}}

\title{Two and three loops computations 
of renormalization constants for lattice QCD\thanks{Presented by V. Miccio}
\vskip-3.6cm\hfill\small HU-EP-04/57; SFB/CPP-04-52; UPRF-2004-19\vskip3.3cm}

\author{F. Di Renzo\address[PARMA]
					{Dipartimento di Fisica, Universit\`a di Parma and INFN, 
				   Gruppo Collegato di Parma, Italy},
 				A. Mantovi\addressmark[PARMA],	
 				V. Miccio\addressmark[PARMA],
        L. Scorzato\address[Humboldt]
        	{Instit\"ut f\"ur Physik, Humboldt Universit\"at, Berlin, Germany},
 				C. Torrero\addressmark[PARMA]
       }

\begin{document}

\begin{abstract}
Renormalization constants can be computed 
by means of Numerical Stochastic Perturbation Theory to two/three loops 
in lattice perturbation theory, both in the quenched approximation and 
in the full (unquenched) theory. As a case of study we report on the 
computation of renormalization constants of the propagator 
for Wilson fermions.
We present our unquenched ($N_f=2$) computations and compare the results with 
non perturbative determinations.
\vspace{1pc}
\end{abstract}

\maketitle

\section*{INTRODUCTION \& MOTIVATION} 

Numerical Stochastic Perturbation Theory (NSPT) is a numerical method \cite{NSPT} to perform Lattice Perturbation calculations. It has been shown to be very powerful in the quenched approximation,
but the technique is quite general and can be in principle applied to any theory \cite{hotQCD}. In particular, it can be generalized to actually carry on unquenched QCD calculations \cite{unquPREV}. The aim is to use NSPT to look at interesting quantities like improvement and renormalization coefficients (quark propagator, currents, etc...), up to 2-3 loops and potentially for any fermionic action. In these pages we will present our results about the critical mass and the field renormalization constant (up to 3 and 2 loops respectively).

\section{COMPUTATIONAL SETUP} 

For a comprehensive account of the strategy of unquenched NSPT we refer the reader to \cite{unquPREV} and references therein.

Simulations are performed on a \emph{APEmille} crate (128 FPU's for 64 GFlops of peak performance), for $N_f=0,2,3$, massless quarks Wilson action on a $32^4$ lattice; gauge is fixed to the Landau condition using a Fourier accelerated algorithm. Stability with respect to single/double precision has been tested, as well as finite size effects are checked looking at smaller lattices on a PC cluster. So far we reached quite a wide configurations database (some hundreds, 1.8GB each) ready for any other observable.

\section{RESULTS}
Both critical mass and field renormalization constant are extracted from measurements of the quark propagator. Actually one averages $S$ on the configurations and then gets $\Gamma_2$ by inversion. The structure of the result is as follow
\begin{eqnarray}
\Gamma_2(p^2,m) &=& S(p^2,m)^{-1} \nonumber \\
&=& i\hspace{-.2em}\not\hspace{-.2em}p - m - \Sigma(p^2,m) 
\end{eqnarray}
where
\begin{eqnarray}
\label{SelfE}
\Sigma(p^2,m) = \Sigma_c + m\:\Sigma_S(p^2,m) + i\hspace{-.2em}\not\hspace{-.2em}p\:\Sigma_V(p^2,m) 
\end{eqnarray}
So, projecting our numerical results onto the \mbox{$\gamma$-matrices} one recovers the last term in the sum and, from it,  one can reach the field renormalization constant; in the same way, projeting onto the \mbox{$\gamma$-identity} one recovers the other two pieces of (\ref{SelfE}), which in (our) massless case reduce just to $\Sigma_c$, the so-called critical mass: it is the additive renormalization for the mass one has to face for having broken chiral symmetry on the lattice.

We present results only for the $N_f=2$ case.

\begin{figure}[t]
	\begin{center}
		\psfrag{p}[ct][ct]{$a^2p^2$}
		\psfrag{0}[rt][rt][.9]{$0$}
		
		\psfrag{0.2}[cc][cb][.9]{$0.2$}
		\psfrag{0.4}[cc][cb][.9]{$0.4$}
		\psfrag{0.6}[cc][cb][.9]{$0.6$}
		\psfrag{0.8}[cc][cb][.9]{$0.8$}
		\psfrag{1}[cc][cb][.9]{$1.0$}
		\psfrag{1.2}[cc][cb][.9]{$1.2$}
		\psfrag{1.4}[cc][cb][.9]{$1.4$}
		\psfrag{1.6}[cc][cb][.9]{$1.6$}
		\psfrag{1.8}[cc][cb][.9]{$1.8$}

		\psfrag{-0.2}[rt][rt][.9]{$$}
		\psfrag{-0.4}[rt][rt][.9]{$-0.4$}
		\psfrag{-0.6}[rt][rt][.9]{$$}
		\psfrag{-0.8}[rt][rt][.9]{$-0.8$}
		\psfrag{-1}[rt][rt][.9]{$$}
		\psfrag{-1.2}[rt][rt][.9]{$-1.2$}
		\psfrag{-1.4}[rt][rt][.9]{$$}
		\psfrag{-1.6}[rt][rt][.9]{$-1.6$}
		\psfrag{-1.8}[rt][rt][.9]{$$}
		
		\includegraphics[scale=0.44]{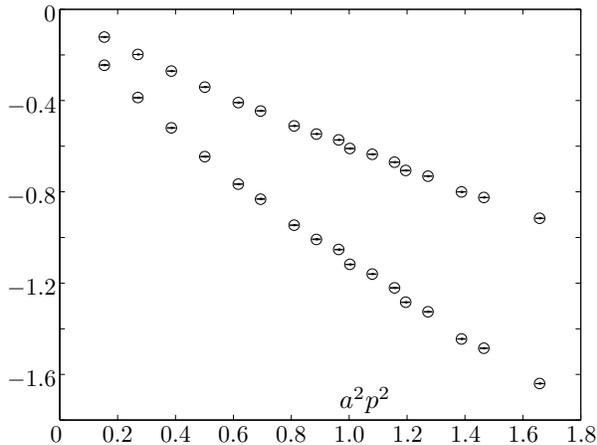}
	\end{center}
	\vspace{-1cm}
	\caption{Renormalized critical mass, $\beta^{-1}$ (upper curve) and $\beta^{-2}$ coefficients; true data are represented by the small points with errorbars; circles represent the same points as they are reconstructed by the fitting procedure.}
	\label{fig:MC1e2}
	\vspace{-.4cm}
\end{figure}

\subsection{Critical mass}

Perturbative expansion for $\Sigma_c$ is already analytically known up to 2 loops \cite{CrMsAnaly1,CrMsAnaly2}. If we write
\begin{equation}
- \Sigma_c = \Sigma_c^{(1)}\beta^{-1} + \Sigma_c^{(2)}\beta^{-2} + \Sigma_c^{(3)}\beta^{-3} + \ldots,
\end{equation}
for $N_f=2$ we have $\Sigma_c^{(1)}=2.6057$ and $\Sigma_c^{(2)}=4.293$.
In order to go beyond these coefficients and get $\Sigma_c^{(3)}$, one must plug them in as counterterms in the simulation because they contribute to higher loops results. So, if one looks at these terms in our results, one actually sees the {\it additive renormalized} coefficients. This is just what is shown in Fig.~\ref{fig:MC1e2}. The non-smoothness of these curves is a consequence of having the continuous euclidean $O(4)$ symmetry broken on the lattice: at finite lattice spacing $\Sigma$ is not simply a function of $a^2p^2$ but it takes also corrections from other lattice invariants one can construct as powers of $ap$. 
Taking into account both the leading $a^2p^2$ ans higher order invariants, one can extrapolate the $a\rightarrow0$ limit value one is interested in. These values are just zero because of the counter\-terms subtraction. 

\begin{figure}[t]
	\begin{center}
		\psfrag{p}[ct][ct]{$a^2p^2$}

		\psfrag{-0.2}[ct][ct][.9]{$$}
		\psfrag{0}[ct][ct][.9]{$0$}
		\psfrag{0.2}[ct][ct][.9]{$$}
		\psfrag{0.4}[ct][ct][.9]{$0.4$}
		\psfrag{0.6}[ct][ct][.9]{$$}
		\psfrag{0.8}[ct][ct][.9]{$0.8$}
		\psfrag{1}[ct][ct][.9]{$$}
		\psfrag{1.2}[ct][ct][.9]{$1.2$}
		\psfrag{1.4}[ct][ct][.9]{$$}
		\psfrag{1.6}[ct][ct][.9]{$1.6$}
		\psfrag{1.8}[ct][ct][.9]{$$}

		\psfrag{7}[r][r][.9]{$7$}
		\psfrag{8}[r][r][.9]{$8$}
		\psfrag{9}[r][r][.9]{$9$}
		\psfrag{10}[r][r][.9]{$10$}
		\psfrag{11}[r][r][.9]{$11$}
		\psfrag{12}[r][r][.9]{$12$}

		\psfrag{6.5}[r][r][.9]{$$}
		\psfrag{7.5}[r][r][.9]{$$}
		\psfrag{8.5}[r][r][.9]{$$}
		\psfrag{9.5}[r][r][.9]{$$}
		\psfrag{10.5}[r][r][.9]{$$}
		\psfrag{11.5}[r][r][.9]{$$}
		
		\includegraphics[scale=0.44]{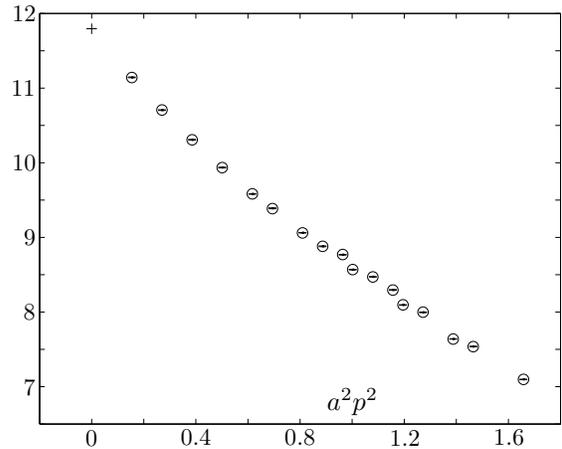}
	\end{center}
	\vspace{-1cm}
	\caption{Critical mass, $\beta^{-3}$ coefficient; circles and points have the same meaning as in Fig.~\ref{fig:MC1e2}; the cross represents the extrapolated value.}
	\label{fig:MC3}
	\vspace{-.5cm}
\end{figure}

Fig.~\ref{fig:MC3} shows the same result, but for the 3-loop term. For it there are no counterterms plugged in the simulations (this is precisely the result one is looking for), and so the extrapolation is not zero. The cross represents the extrapolated $(a\rightarrow0)$ value, which reads
\begin{equation}
\Sigma_c^{(3)} = 11.79^{(2)}_{(5)} 
\end{equation}
in which the errors are estimated by looking at the stability of the fitted result with respect to varying the number of points included in the fit and the higher orders lattice invariants taken into account.

\subsection{Field renormalization constant}
As said, the field renormalization constant $Z_q$ can be extracted looking at $S^{-1}$ along the \mbox{$\gamma$-matrices}
\begin{eqnarray}
Z_q&=&-\frac{i}{4}\frac{{\rm Tr}(\hspace{.1em}\not\hspace{-.2em}p\hspace{.2em}S^{-1})}{p^2} \nonumber \\ 
&=& 1 + Z_q^{(1)}\beta^{-1} +  Z_q^{(2)}\beta^{-2} + \ldots
\end{eqnarray}
In contrast with the case of the critical mass, in this case we have to face logarithmic divergences, $\log(a^2p^2)$, coming from Renormalization Group Theory. These divergences are ruled by anomalous dimensions and so, because we are in Landau gauge, we have just {\sl no} \mbox{$\log$-terms} at \mbox{1-loop} and only a {\sl simple} \mbox{$\log$-term} at \mbox{2-loops} (i.e. without \mbox{$\log^2$-term}).

\begin{figure}[t]
	\begin{center}
		\psfrag{p}[ct][ct]{$a^2p^2$}

		\psfrag{0}[ct][ct][.9]{$0$}
		\psfrag{0.2}[ct][ct][.9]{$0.2$}
		\psfrag{0.4}[ct][ct][.9]{$0.4$}
		\psfrag{0.6}[ct][ct][.9]{$0.6$}
		\psfrag{0.8}[ct][ct][.9]{$0.8$}
		\psfrag{1}[ct][ct][.9]{$1.0$}
		\psfrag{1.2}[ct][ct][.9]{$1.2$}

		\psfrag{0.75}[r][r][.9]{$0.75$}
		\psfrag{0.77}[r][r][.9]{$0.77$}
		\psfrag{0.79}[r][r][.9]{$0.79$}
		\psfrag{0.81}[r][r][.9]{$0.81$}
		\psfrag{0.83}[r][r][.9]{$0.83$}
		\psfrag{0.85}[r][r][.9]{$0.85$}
		
		\includegraphics[scale=0.44]{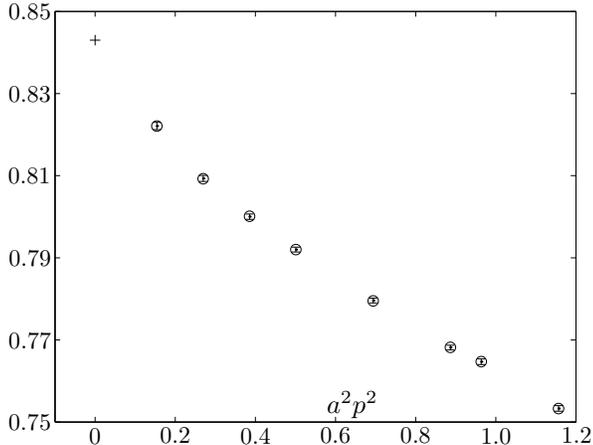}
	\end{center}
	\vspace{-1.1cm}
	\caption{Field renormalization constant, $\beta^{-1}$ coefficient; marks have the same meanings as in Fig.~\ref{fig:MC3}.}
	\label{fig:Zq1}
	\vspace{-.5cm}
\end{figure}

So, at 1-loop one has no divergences to subtract and one can extrapolate directly to the $a\rightarrow0$ limit. The result we found is shown in Fig.~\ref{fig:Zq1} and reads
\begin{equation}
Z_q^{(1)}=-0.843^{(2)}_{(7)}.
\end{equation}
This result is in perfect agreement with the known value $Z_q^{(1)}=-0.843...$

\begin{figure}[t]
	\begin{center}
		\psfrag{p}[ct][ct]{$a^2p^2$}

		\psfrag{0.0}[ct][ct][.9]{$0.0$}
		\psfrag{0.1}[ct][ct][.9]{$$}
		\psfrag{0.2}[ct][ct][.9]{$0.2$}
		\psfrag{0.3}[ct][ct][.9]{$$}
		\psfrag{0.4}[ct][ct][.9]{$0.4$}
		\psfrag{0.5}[ct][ct][.9]{$$}
		\psfrag{0.6}[ct][ct][.9]{$0.6$}
		\psfrag{0.7}[ct][ct][.9]{$$}
		\psfrag{0.8}[ct][ct][.9]{$0.8$}
		\psfrag{0.9}[ct][ct][.9]{$$}
		\psfrag{1.0}[ct][ct][.9]{$1.0$}

		\psfrag{1.16}[r][r][.9]{$1.16$}
		\psfrag{1.18}[r][r][.9]{$$}
		\psfrag{1.2}[r][r][.9]{$1.20$}
		\psfrag{1.22}[r][r][.9]{$$}
		\psfrag{1.24}[r][r][.9]{$1.24$}
		\psfrag{1.26}[r][r][.9]{$$}
		\psfrag{1.28}[r][r][.9]{$1.28$}
		\psfrag{1.3}[r][r][.9]{$$}
		\psfrag{1.32}[r][r][.9]{$1.32$}
		\psfrag{1.34}[r][r][.9]{$$}
		\psfrag{1.36}[r][r][.9]{$1.36$}

		\includegraphics[scale=0.44]{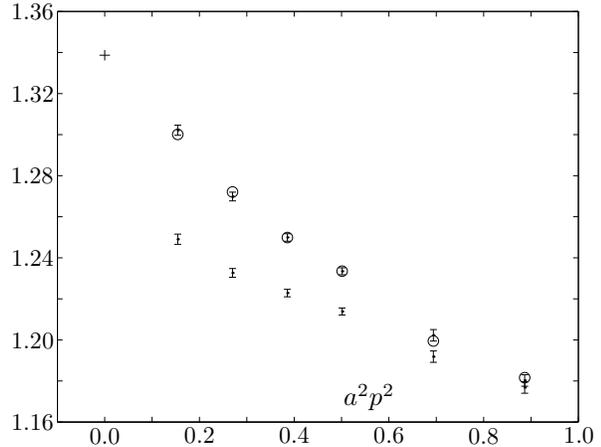}
	\end{center}
	\vspace{-1.1cm}
	\caption{Field renormalization constant, $\beta^{-2}$ coefficient; marks have the same meanings as in Fig.~\ref{fig:MC3}; crude data are on the lower curve; on the upper curve the log-divergence is subtracted.}
	\label{fig:Zq2}
	\vspace{-.5cm}
\end{figure}

At 2-loops one has to subtract the known \mbox{log-term} before taking the continuum limit. In Fig.~\ref{fig:Zq2} we plot both the crude data and the data with the log divergence subtracted: the extrapolation for the latter reads
\begin{equation}
Z_q^{(2)}=-1.34(2) .
\end{equation}
We can compare this new result with a non-perturbative value \cite{Cecilia}, which at $\beta=5.8$ and $a\mu=1$ reads $Z_q^{\rm(N\!-\!P)} = 0.786(5)$. Our \mbox{2-loop} expansion add up to $Z_q^{\rm(P)} = 0.815$. So the discrepancy reduces from 9\% \mbox{(1-loop)} to \mbox{3-4\%}. In order to improve convergence, one can switch from a bare to a better-converging coupling expansion (such as in the so-called Boosted Perturbation Theory). On the other hand we already have the signal for the \mbox{3-loop} coefficient.

\section{CONCLUSION \& PERSPECTIVES}
We are currently refining the statistic in order to reduce statistical errors and to see the \mbox{3-loop} coefficient for $Z_q$.

{\sl Just} the same strategy used here is needed to measure fermion currents and the renormalization constants related to them (and we are actually working at it) as well as to investigate the twist-mass QCD action.

{\sl Much} the same strategy can also give improvement coefficients.

{\sl Almost} the same strategy may also allow to perform (quenched) overlap fermion calculations (and we are actually working on it as well).


\begin{thebibliography}{1}

\bibitem{NSPT}
F. Di Renzo, G. Marchesini and E. Onofri, P. Marenzoni, Nucl. Phys. B426 (1994) 675
\bibitem{hotQCD}
F. Di Renzo, A. Mantovi, V. Miccio, Y. Schroder, JHEP 0405:006,2004
\bibitem{unquPREV}
F. Di Renzo and L. Scorzato, Nucl. Phys. B Proc. Suppl. 94 (2001) 567
\bibitem{CrMsAnaly1}
E. Follana, H. Panagopoulos, Phys. Rev. D 63 (2001) 17501
\bibitem{CrMsAnaly2}
S. Caracciolo, A. Pelissetto, A. Rago, Phys. Rev. D 64 (2001) 94506
\bibitem{Cecilia}
V. Lubicz, C. Tarantino, private communication. See their contribution at this conference.



\end{thebibliography}
\end{document}